# Overview of Optical Interconnect Technology

Sumita Mishra, Naresh K Chaudhary, Kalyan Singh

**Abstract**— Optical interconnect is seen as a potential solution to meet the performance requirements of current and future generation of data processors. Optical interconnects have negligible frequency dependent loss, low cross talk and high band width. Optical interconnects are not much used commercially since optical interconnects technology is incompatible with manufacturing processes and assembly methods that are currently used in the semiconductor industry. There are many promising optical interconnect technologies and this paper presents a brief analysis of current state of optical interconnect technology.

——————————  ◆  ——————————

## 1 INTRODUCTION

In recent years, the performance of single chip multiprocessor has been roughly doubling every 18 months resulting in tremendous increase in its clock speed and cache size moreover multiple cores and hyper threading has increased the complexity of these systems.

These multicore ultrafast processors require high speed interconnects that allow individual processors fast access to memory, its cores and other I/O devices. As speed and complexity of these systems increase the interconnect density and throughput management becomes a critical factor towards the realization of high performance data processing systems. Currently copper interconnects are used for data transmission over chip-to-chip and chip-to-module interfaces, in chip-to-chip over backplane, and in chip-to-chip over copper cable assemblies. Electrical interconnects performance degrades at frequencies above 1 GHz due to ringing, increased signal latency, crosstalk and frequency dependent attenuation [1]. These limitations of electrical interconnects will limit the maximum frequency of operation for future systems; Optical interconnect is seen as a potential solution since it can directly address these problems at the system level and meet the performance requirements of current and future generation of data processors, optical interconnects have negligible frequency dependent loss, low cross talk and high band width.

Despite the significant interest shown by many groups worldwide, optical interconnects are not much used commercially. In order to become a viable technology to replace electrical-based on-chip interconnects, optical interconnects should be made compatible with manufacturing processes and assembly methods that are already in use in the semiconductor industry further there is a need to develop efficient and compact optical interconnect modules that use simple optical and electrical interfacing schemes. The development of optical interconnects, especially based on a technology platform which is monolithically integratable into Si CMOS at low cost is needed in order to make optical interconnects economically viable. This will result in low cost, high performance and CMOS compatible optical components. Since it is not possible to make silicon light emitting and detecting we have to integrate other materials with Si. Large-scale integration of optical devices has been demonstrated on III-V platforms but in this implementation the components have different technology and they cannot be monolithically integrated on the same substrate. Monolithic integration of optical and electronic components on one substrate [2-7] together with demonstrations of efficient fibre to waveguide couplers [8] has shown the promise for development of ultra-compact optical components compatible with current technology.

## 2 BASIC COMPONENTS

The main components of an optical interconnect system are shown in Figure 1. The optical modulator has two inputs, optical signal from the off chip laser source and the electrical signal from the CMOS driver circuit. Optical couplers are structures that are used to inject the light into the optical system. Electrical signals that are to be transmitted to some destination in an optical interconnect system must be converted into the

————————
*Sumita Mishra is currently pursing doctoral degree in Electronics at DRML Awadh University, India*
*E mail: mishra.sumita@gmail.com*

optical domain for transmission. Modulator converts the electrical signal into optical signal according to bit sequence in electrical signal .After the optical information signal has been generated, it is fed into the optical routing structure. Optical interconnects use waveguides for signal transmission, which consists of dielectric materials with high index of refraction surrounded by a material with lower refractive index. Optical switches are used in optical routing networks to route the light travelling in waveguides to different locations.

The receiver side of the optical interconnect system is responsible for reconstruction of electrical signal. A photo detector is the device for detecting the light pulses and converting them to photo current. A trans-impedance amplifier is finally used for amplifying the photo current and providing the digital signal in the form of conventional voltage signal.

GaAs/AlGaAs-based process technology. These devices have advantage of emission through the substrate, since GaAs is transparent at 980 nm. This brings added benefits such as integrating backside micro lenses for improved coupling [10]. 980 nm VCSELs have been the wavelength of choice for an ultra-dense chip-to-chip level interconnects, operating at data rates in the range of Tb/s . During the recent years there has been a significant development effort towards VCSELs operating at a wavelength range of 1200…1600 nm using the InP-based process technology. VCSELs operating at a wavelength near 1550 nm have the disadvantage of growth difficulties and low thermal conductivity of the necessary DBRs. [11] VCSELs offer many of the desirable characteristics of an optoelectronic transmitter for optical interconnects. 2D arrays of high speed VCSEL with good contrast at a low voltage drive can be fabricated . Recent advances in VCSEL technology [12]along with research into new materials such as InGaNAs(Sb) has shown promise for higher data rates with low power consumption in the near future

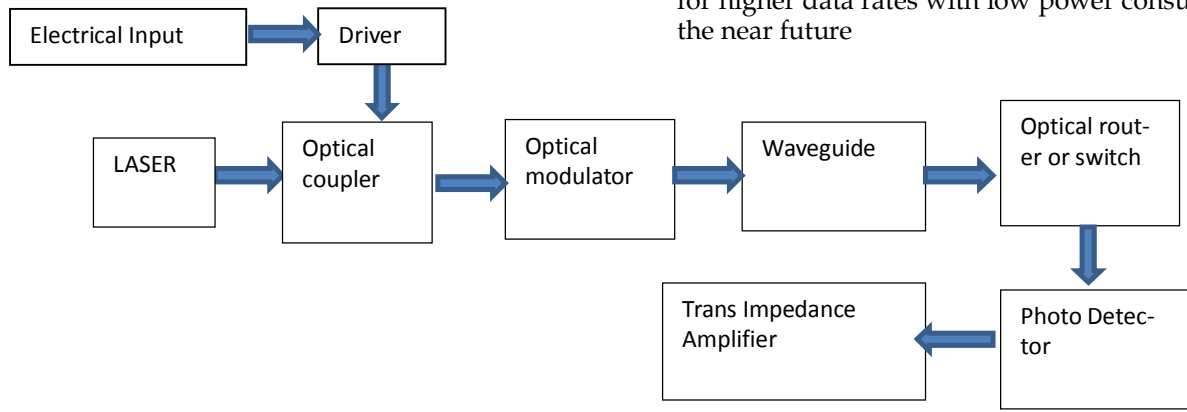

**Fig 1.** Optical interconnect block diagram

## 2.1 Laser source

Vertical Cavity Surface Emitting Laser (VCSEL) is currently the most attractive optical source for the short-distance optical interconnects. A Vertical Cavity Surface-Emitting Laser (VCSEL) is a semiconductor laser diode that emits light perpendicular to the upper surface of the semiconductor wafer of which the laser is composed. Monolithic arrays of high density VCSELs [ref] have been developed by various manufacturers which are competitively priced. VCSELs can be manufactured for several different wavelengths [9].Today, devices emitting at around 850 nm represent the most mature high speed-optimised VCSEL technology these devices are top emitting due to large substrate. VCSELs operating at 980 nm are based on

## 2.2 Optical modulator

Optical interconnect systems may use either off-chip or on-chip laser sources. Off-chip laser sources are generally used due to several problems faced by directly modulating Lasers. Due to non-availability of an efficient silicon based laser it is very hard to fabricate large no of lasers on a single chip at reduced cost further it has added disadvantage of complex chip design since optical source is part of the chip's power and heat budget. In order to modulate lasers at high bit rates, they must be operated well above threshold in both 0 and 1 bit states, which consumes high power and leads to performance degradation. One of the critical issues of future systems is to keep power budgets manageable while increasing performance this is achieved by increasing parallelism. Wavelength division multiplexing offers massive parallelism. WDM components are not an essential part for optical interconnect systems but they are considered very important in order to build high performance optical interconnects. The heat generation from on chip lasers is undesired since the temperature variation in CMOS chips also causes wavelength shifts and this instability can prohibit precise channel allocations for wavelength Division multiplexing in the same medium. So it is preferable to use on-chip modulators for transmitters and modulate the light coming from an off-chip continuous-wave (CW) laser. The design of a fast and cost efficient CMOS compatible electro-optical modulator is one of the most challenging tasks towards realizing on-chip optical interconnects.

There are various optical modulation techniques through which refractive index or absorption properties of optical medium are varied in accordance with the electrical signal; this variation in optical property of medium causes phase or amplitude modulation of optical signal . Current optical modulation techniques are based on Thermo Optic effect, Electro optic effect,[13] Electro absorption effect and plasma dispersion effect. Thermo-optic effect is the change in optical absorption coefficient and refractive index of an optical medium due to the change in temperature of the medium. Refractive index variation due to applied electric field is called Electro Optic Effect. The linear change in a optical medium's refractive index by an electric field is called the Pockels effect and if the change is quadratic, it is called the Kerr effect. Another modulation mechanism known as the electroabsorption effect involves the change in the absorption coefficient of the material with change in applied electric field. The Franz–Keldysh effect [14] is electroabsorption observed in bulk semiconductors and the Quantum Confined Stark Effect [15] is electroabsorption seen in quantum-confined structures such as thin quantum-well layers.

These modulation techniques are quite effective in III – IV semiconductors but in silicon refractive index change produced by these effects is very small since unstrained pure crystalline silicon does not exhibit Pockel's effect, Franz–Keldysh effect and the Kerr effect are very weak in Si Therefore; very high electromagnetic field is required in order to achieve a useful change in the refractive index. The most effective mechanism for changing the refractive index in Si is the carrier plasma dispersion effect [16] in which the concentration of free carriers in silicon changes the refractive index and the optical absorption.

The modulator can be either refractive or absorptive type. Refractive modulators employ either single interference Mach–Zehnder interferometer structure or multiple interference resonator structure . In Mach Zehnder interferometer two light beams pass through the two arms of the MZ structure, refractive index in one of the arms is varied through some modulation mechanism, resulting in different optical path lengths. The superposition of the two beams at the other end will result in modulation of optical signal. [17]. Modulators using MZ structure are large in size and have slow transition time.

The use of resonator structure enables modulation using compact devices. The transmittivity of a resonator is given approximately by

$$T = \frac{T_{max}}{1+\left(\frac{4nL}{\pi \Delta \lambda}\right)^2 Sin^2\left(\frac{2\pi nL}{\lambda}\right)}$$

where $\lambda$ is the wavelength of light incident on the resonator, and $\Delta\lambda$ is the full-width at half maximum of the transmission at resonance. The resonator circulates light within the cavity at the resonance wavelength which increases the optical path length without increasing the physical device length. If the incident wavelength is resonance wavelength light exits the device with transmittivity T = Tmax after a photon life time. At all other wavelengths light destructively interferes in the cavity and is not transmitted through the resonator. From equation 1 it is observed that if the refractive index is changed the transmission changes, resulting in amplitude modulation of the signal transmitted through the resonator.

Absorptive modulators based on electroabsorption effect work by changing the optical absorption in the modulator structure by application of electric field which is obtained by reverse biasing PIN diodes that contain the bulk semiconductor or quantum well materials in the intrinsic region of the diodes. Many Quantum-well modulator devices have been demonstrated at high speeds employing only a few micrometers of optical path length even without the use of resonators [18-20], Although use of cavities can enhance the performance of these devices . Another approach is to bond III–V devices to silicon in a waveguide configuration. Resulting in InAlGaAs QCSE modulators bonded to silicon structures[21].

Recently electroabsorption modulators employing Ge quantum well structures grown on silicon has been demonstrated[21]. A lot of research is being done on optimizing device structures and developing integration techniques for these Si and Ge based high speed and low power modulators [21-29]which are compatible with current CMOS technology.

## 2.3 Receiver

Receiver section in an optical interconnect comprises of a semiconductor Photo Detector followed by an electronic amplifier. The photo detector performs the operation of optical-to-electrical signal conversion. It is basically a reverse biased device which absorbs the incident radiation and generates electron-hole pairs which in turn produces a photo current in the external circuit . The detector to be used in optical interconnects should provide high bandwidth, high sensitivity and easy optical coupling, moreover it should be amenable to high density fabrication. There are many semiconductor photo detectors that may fulfil these requirements: p-i-n photodiodes (PIN), metal-semiconductor-metal photodiodes (MSM), and avalanche photodiodes (APD). PINs are the most commonly used photo detector in short-distance optical links. MSM is a low-capacitance optical detector with high-speed operation and a larger active area than PIN but the responsivity of MSM is typically lower than of PIN and although APD provides the highest responsivity due to the internal gain, its drawbacks are the requirement of high-bias voltage and expensive fabrication process.

GaAs is a suitable material for photodiodes at wavelengths up to 850 nm and GaAs is more popular in high-speed applications but it is expensive. Si is used at wavelengths up to 1 μm; In longer wavelengths up to 1.7 μm, an InGaAs material system on InP substrates is commonly used, but for optical interconnects the most important detector material is Ge, because of its potential to monolithically integrate photo diodes on Si-based integrated circuits. For optical interconnects waveguides are used as routing device thus, instead of being surface-normal, the Photo Diode needs to be waveguide integrated. Recently Waveguide-integrated Ge photo diodes have become a topic of intense research and development [30-41].

The small photocurrent generated by the Photo Diode must be amplified with a minimum amount of added noise for further processing. Hence, a preamplifier is used as the first stage of amplification. Three widely used configurations for pre amplifier circuits are the high input impedance amplifier, the low input-impedance amplifier, and the trans impedance amplifier. TIA is typically chosen in optical interconnects because it meets the requirement of large bandwidth at low noise.

## 2.4 Wave guides, Switches, Routers

Optical waveguide to be used in optical interconnects must have a very low attenuation and good optical properties with high stability against stresses involved in the electronics assembly processes. The attenuation in silicon waveguides arises mainly due to light scattering from the etched sidewalls. Minimizing the optical field overlap with etched interfaces can effectively reduce the attenuation in waveguide. Increasing waveguide width and decreasing etch depth minimizes this overlap. Hence ridge waveguide structure is generally preferred for waveguides in optical interconnects. Several waveguide manufacturing technologies exist such as photolithography, printing, ion-exchange, laser direct writing, and laser ablation. The materials used for fabrication of waveguides include acrylates, polyimides, cyclic olefins, siloxanes and silsesquioxanes. [43-49]. Apart from waveguide other passive waveguide components, such as bends, splitters and crossings are essential for signal routing on the board.

## 3. CONCLUSIONS

The motivation for using optical interconnects technology is that the electrical interconnects cannot keep track of Moore's law indefinitely with increasing data rate requirements of high performance computers back frame computers , supercomputers and other high speed data processors . One possible alternative is use of optical interconnects. The main advantages offered by optical interconnects over their electrical counterparts are higher carrier frequency, less attenuation, less crosstalk and lower power consumption. Despite these advantages, there are a number of issues to be resolved before optical interconnect technology can be implemented.

Currently there are number of optical interconnect technologies at various stages of development. III-V quantum wells are considered the most mature technology which is already in the market. High index contrast structures are also being integrated in various chip configurations. The rest of the technologies are still in either fundamental research or applied research phase and hence are not ready for the market. A lot of research and is still required for enhancing their performance. The incompatibility of mature III-V semiconductor materials with current CMOS technology is a major hurdle in economic production of reliable integrated circuits employing optical interconnects. Integration of components and manufacturing integrated circuits using optical interconnects requires significant research investments. The fact remains that silicon based technology which is compatible with current integrated circuit technology still need to be proven efficient technology for optical interconnects. The performance and reliability of available optical components needs to be improved significantly, moreover electrical interconnect technology is still good enough thus for most applications, optical interconnects are used for links requiring the highest data rates.

## ACKNOWLEDGMENT

The first author Sumita Mishra is grateful to Maj. Gen. K.K. Ohri , Prof. S.T.H. Abidi and Brig. U. K. Chopra of Amity University, India for their support during the research work.